\documentstyle[prl,aps,twocolumn]{revtex}
\tightenlines
\begin{document}
\draft
\title{Temperature-frequency scaling in amorphous niobium-silicon near the
metal-insulator transition}
\author{Hok-Ling Lee, John P. Carini, and David V. Baxter}
\address{Department of Physics, Indiana University, Bloomington, IN 47405}
\author{George Gr\"uner}

\address{Department of Physics, University of California, Los Angeles, CA
90024}

\date{\today}
\maketitle
\begin{abstract}
Millimeter-wave transmission measurements have been performed in amorphous
niobium-silicon alloy samples where the DC conductivity follows the critical
temperature dependence $\sigma_{dc} \propto T^{1/2}$.
The real part of the conductivity is
obtained at eight frequencies in the range 87--1040 GHz for temperatures 2.6 K
and above.  In the quantum regime ($\hbar \omega > k_B T$) the real part of the
high-frequency conductivity has a power-law frequency dependence ${\rm
Re}~\sigma(\omega) \propto \omega^{1/2}$.  For temperatures 16 K and below the
data exhibits temperature-frequency scaling predicted by theories of dynamics
near quantum-critical points.
\end{abstract}
\pacs{72.15.Rh,72.30.+q,72.60.+g}

The interplay between dynamics and statistical mechanics in quantum phase
transitions suggests that experiments probing the frequency
dependence of such systems will be useful in understanding the physics of
quantum-critical points, where quantum fluctuations have diverging
characteristic length and time scales, $\xi$ and $\xi_{\tau}$ respectively
\cite{sachdev97,damle97,sondhi97}.  For example, recent studies of the
quantum-Hall effect \cite{engel93,engel95} and two-dimensional
superconductor-insulator transition \cite{yazdani95} have exploited this
approach.  The disorder-induced metal insulator transition in bulk systems has
been studied for decades as a quantum-critical point \cite{anderson58,wegner76}
(and see \cite{belitz94} for a recent review), yet few experiments have studied
the frequency-dependent conductivity in the quantum regime, $\hbar \omega > k_B
T, \hbar/\xi_{\tau}$ (exceptions being far-infrared conductivity
measurements in Si:P near its metal-insulator transition
\cite{rosenbaum81,gaymann93}) or in the crossover regime where $\hbar \omega
\simeq k_B T$.

Even without a microscopic model for the transition, several general
predictions can be made on the basis of scaling arguments for the critical
behavior of the conductivity as a function of temperature and frequency,
$\sigma_c(T, \omega)$.  For $\hbar \omega \gg k_B T $, one directly observes
the quantum fluctuation conductivity, $\sigma_c \propto \omega^{1/z}$ (where
$z$ is the dynamical exponent); for $k_B T \gg \hbar \omega$, one expects
\cite{sondhi97} the thermal energy scale to limit the duration of quantum
fluctuations resulting in a DC conductivity varying as $\sigma_c \propto
T^{1/z}$.  For intermediate temperatures and frequencies, a universal crossover
function that depends only on the ratio $\frac{\hbar \omega}{k_B T}$
interpolates between the limiting behaviors.

We report measurements of the millimeter-wave transmission through an amorphous
niobium-silicon alloy sample as a function of temperature.  This sample has
a temperature dependence for its DC electrical conductivity indicating that the
niobium concentration is very close to the critical value that divides alloys
having zero DC electrical conductivity for $T \rightarrow 0$ (insulators)
from those which conduct for $T \rightarrow 0$.
The experiment spans a sufficiently broad range of frequency that it probes
the quantum regime for the
lowest experimental temperatures.  The transmission data are analyzed to obtain
the magnitude of the real part of the conductivity as a function of temperature
and frequency.  In the quantum regime, the data is consistent with a power-law
frequency dependence for the conductivity ${\rm Re}~\sigma \propto
\omega^{1/2}$.  For a range of temperatures from 2.6 K to 16 K we find that the
temperature and frequency dependances of ${\rm Re}~\sigma$ are consistent with
the existence of a crossover function.

Amorphous niobium-silicon samples are grown onto 0.5 mm thick sapphire
substrates using a sequential sputtering technique.  The substrates are mounted
on a rotating stage and are exposed alternately to pure silicon and niobium
sources, which deposit bilayers of approximately 1 nm amorphous silicon and a
sub-monolayer of niobium.  This technique produces large (15 mm diameter),
homogenous samples with thickness up to 0.5 $\mu$m, which are needed
for the optical measurements.  The rotational rate of the substrate over each
source determines the niobium concentration of each sample.  Low angle x-ray
diffraction experiments demonstrate near-total interdiffusion of the niobium
and silicon, consistent with previous reports of niobium-silicon multilayers
\cite{song91}.  In addition to this large sample, a sample suitable for dc
electrical transport experiments is grown simultaneously in a Hall-bar
configuration on an adjacent sapphire substrate.

The temperature dependence of the dc conductivity for our samples is very
similar to that found near the critical niobium concentration in other studies
of amorphous Nb-Si \cite{hertel83,bishop85,allen93,carini96}, which all found a
temperature dependence for the DC critical conductivity $T^{1/2}$, implying
$z=2$.  For the sample used in the transmission measurement,  the temperature
dependence of the DC conductivity in the temperature range 1.4--15 K is $\sigma
= \sigma_0 + 475~T^{1/2}(\Omega {\rm m K}^{1/2})^{-1}$, where $\sigma_0 =
125~(\Omega m)^{-1}$ (the absolute uncertainty is 5\%).  The positive, small
value for $\sigma_0$ implies that this sample is barely metallic and has a
characteristic quantum energy scale that is small enough so that experiments
for temperatures above 1~K (or frequencies with $\frac{\hbar \omega}{k_B} \gg
1~{\rm K}$) are probing the quantum-critical behavior of this alloy system
\cite{sondhi97}.  Tunneling experiments support the existence of a small
characteristic quantum energy scale in similar samples
\cite{hertel83,bishop85}.  The conductivity for this sample begins rising above
the $T^{1/2}$ dependence for $T>15~{\rm K}$.

The millimeter-wave conductivity measurements are performed by placing a sample
as one window of a movable holder, which also contains an empty window, in the
tail of an optical cryostat.  Several different backward-wave oscillator
sources produce a few milliwatts of tunable monochromatic radiation in the
frequency range 75--1080~GHz.  During the experiment, the frequency is dithered
by a few percent in order to average over interference oscillations in the
transmission produced by multiple reflections between other objects in the
optical path.  Because the sapphire substrates are birefringent, a wire-grid in
front of the sample polarizes the incident radiation along the sapphire
$c$-axis (the higher-index axis).  The transmission (the ratio of the power
transmitted through the sample to that transmitted through the empty window) is
measured as a function of frequency and temperature for both the sample and a
blank substrate (for the same relative polarization as for the sample).

The transmission oscillates as a function of frequency because the substrate
acts like a Fabry-Perot resonator with a rather low Q-factor (see the inset of
Figure 1, the transmission of the sample at two temperatures as a function of
frequency and for a substrate without a sample at room temperature).  In
principle, if the thickness $d_1$ and refractive index $n_1$ of the substrate
and the thickness of the sample $d$ are known, then the complex conductivity of
the film $\sigma(\omega)$ can be determined at a discrete set of frequencies
from the magnitude of the transmission at each peak frequency
$|t(\omega_{peak})|^2$ and the shifts of the peak frequency value.  In
practice, experimental uncertainties make this inversion problematic; however
in the limit that $(|\sigma| d Z_0)/n_1 \ll 1$ (where $Z_0 \approx 377~\Omega$
is the impedance for electromagnetic waves in vacuum) one finds this
approximate linear relationship (accurate to 3\% or better in our experiment)
from standard transmission-line theory:
\begin{equation}
{\rm Re}~\sigma(\omega_{\rm{peak}}) \approx \frac{2}{Z_0 d}
\left(\frac{1}{|t(\omega_{\rm{peak}})|} - 1\right).
\end{equation}
This relation assumes no absorption in the substrate; as shown by the data for
the substrate in the inset, absorption by the substrate alone becomes
noticeable for the higher measurement frequencies, but correcting for its
presence is straightforward.

The main panel in Figure 1 shows the peak transmission (corrected for
absorption by the substrate) as a function of temperature for eight peak
frequencies ranging from 87--1040 GHz.  At low temperature, the transmission
is much closer to unity for the lower frequencies than the higher frequencies,
which shows that ${\rm Re}~\sigma$ increases strongly with frequency in this
frequency range.  At higher temperature, the transmission is lower for all
frequencies than at lower temperature (indicating that ${\rm Re}~\sigma$ is
greater for high temperatures) and the transmission values are
approximately the same for the different frequencies, indicating that the
conductivity is essentially independent of frequency at this temperature.

Figure 2 shows the frequency-dependence of the real part of the conductivity,
plotted as a function of the square-root of frequency, for several temperatures
in the range 2.8--32 K.  For the lowest temperatures (2.8 K and 4.7 K, where
$k_B T < \hbar \omega$ except for the lowest frequency), the data follow a pure
$\omega^{1/2}$ frequency dependence (as indicated by the straight line in the
figure). For the intermediate temperatures (where $k_B T$ falls in the middle
of the range of $\hbar \omega$), the data approach a similar $\omega^{1/2}$
dependence only for the highest frequencies and approach the dc values for
lower frequencies.  For 40 K (where $k_B T > \hbar \omega$) the increase in the
conductivity with increasing frequency has vanished.

The systematic crossover from a frequency-determined to a
temperature-determined conductivity when $k_B T \simeq \hbar \omega$ suggests
the existence of a crossover function $\Sigma$ of the form expected for
quantum-critical dynamics \cite{sachdev97,damle97}:
\begin{equation}
\sigma(x_c, T, \omega) = C T^{1/z} \Sigma\left(\frac{\hbar \omega}{k_B
T}\right).
\end{equation}
One can test for the existence of such a function using the dynamical exponent
value $z=2$ and $C=475 (\Omega {\rm m} K^{1/2})^{-1}$ (as suggested by the
temperature dependence of the dc conductivity data at the critical point) and
plotting the scaled conductivity data ${\rm Re}~\sigma(T, \omega)/(C T^{1/2})$
versus scaled frequency $\frac{\hbar \omega}{k_B T}$; the results are shown in
Figure 3.  The scaling procedure collapses the data over the entire frequency
range for temperatures 16 K and below, indicating the existence of a single
crossover function that depends only on the scaled frequency.  Higher
temperature data begins to rise above the collapsed data especially for lower
frequencies.  Varying the value of $z$ by more than 10\% makes the collapse
much less satisfactory.

The curve defined by the collapsed data has several features that a
quantum-critical crossover function should possess.  The demarcation between
the flat part of the curve ($T$-dominated dynamics) and the increasing part of
the curve ($\omega$-dominated dynamics) occurs for $\frac{\hbar \omega}{k_B T}
\simeq 1$.  This indicates that the thermal energy defines the time-scale
$\frac{\hbar}{k_B T}$ that limits the size of the quantum fluctuations (as
measured by the frequency dependent conductivity), which is a key prediction of
quantum-critical scaling theory.  In addition, for high scaled frequencies, the
crossover function follows a power-law dependence on scaled frequency with the
same value for the dynamical exponent as was used to scale the
vertical axis in the Figure.  This means that one could have found a crossover
function by scaling the conductivity data by $\omega^{1/2}$ and plotting the
result as a function of $\frac{k_B T}{\hbar \omega}$.

An attempt to guess the form of the universal crossover function, $\Sigma =
{\rm Re}~(1 - \imath b \hbar \omega/(k_B T))^{1/2}$ where $b$ is supposed to be
a dimensionless constant of order unity, is shown by the dashed line in Figure
3.  This function describes the high and low frequency behavior of the
collapsed data but misses the sharp crossover in the data for $\hbar
\omega/(k_B T) \simeq 1$.  This expression implies that there should be a
sizeable imaginary component for the conductivity, which is negative in sign
and, for exponent 1/2, equal in size to the real part for high frequencies.
The imaginary part of the conductivity can be extracted in this type of
experiment from the shifts of the peak transmission frequencies; however these
shifts turn out to be small and difficult to measure accurately when the index
of refraction of the substrate is large (as it is here for sapphire, 3.35).
For the four low frequency peaks (85--340~GHz), we do observe peak shifts as a
function of temperature that have the correct sign and the expected magnitude.

The fact that the scaling works only for temperatures 16 K and below is
reasonable given the behavior of the dc conductivity for the sample, which
begins to deviate from a $T^{1/2}$ temperature dependence for higher
temperatures.  This implies that there is an additional mechanism that becomes
active for higher temperatures (electron-phonon interactions, for example) and
a new time scale (in addition to $\frac{\hbar}{k_B T}$) that acts to limit the
size of quantum fluctuations.

The value of the dynamic exponent $z=2$ is consistent with previous work on
this alloy system close to the metal-insulator transition, including the
temperature dependence of the dc conductivity \cite{bishop85,allen93} and
tunneling measurements \cite{hertel83}.  This value is different from what is
expected from interacting models without a density of states singularity at the
transition ($z=d$ the spatial dimension) or those with straight Coulomb
interactions ($z=1$).  Models with a screened Coulomb interaction can produce
$z=2$ \cite{macmillan81,efros85}.  Field-theoretical treatments also find
several scenarios where the value $z=2$ is possible \cite{belitz94}.

At least one other quantum-critical point has been investigated using combined
frequency- and temperature- dependent conductivity experiments in the quantum
regime, the magnetic field tuned transition between different integer
quantum-Hall states \cite{engel93,engel95}.  Here the crucial quantity was not
the conductivity itself, since in a two-dimensional system the critical
conductivity approaches a constant of order $e^2/h$.  Instead, the widths (as a
function of magnetic field) of the peaks in the real part of the conductivity,
which occur between the plateaux in the Hall resistance, were measured as a
function of frequency and temperature.  Analysis of the widths leads to the
temperature-frequency scaling near the critical field magnetic value and a
collapse of the data similar to what we observe. The dynamical
exponent was found to be $z=1$.  The crossover from temperature-dominated to
frequency-dominated dynamics at the critical magnetic field appeared to occur
for $\frac{\hbar \omega}{k_B T} \simeq \frac{1}{3} $ \cite{engel95}, which is
somewhat different from the crossover location we observe.

In conclusion, our experiments on the temperature and frequency dependances of
the conductivity of an amorphous niobium-silicon sample are consistent with the
quantum-critical picture of the dynamics.  In particular,
from frequency-dependent experiments on a near-critical sample we find a
crossover between a high-frequency regime dominated by quantum fluctuations and
a low-frequency regime dominated by thermal fluctuations.  The crossover itself
occurs for $\frac{\hbar \omega}{k_B T} \simeq 1$, which implies that the
thermal energy directly limits the duration of the quantum fluctuations in the
quantum-critical state.  The form of the crossover function itself, in
particular the location of the crossover and the dynamical exponent, have not
yet been explained by theory.

Acknowledgements---We thank A. Schwartz and C. Hillman for assisting with the
transmission experiments and D. Belitz, R. Bhatt, S. Girvin, T. Kirkpatrick,
and S. Sachdev for fruitful conversations.  JPC thanks the Aspen Center for
Physics.  This work was supported by NSF grants DMR-9314018, DMR-9423088, and
DMR-9503009.


%
%

\begin{figure}
\caption{Peak transmission (corrected for absorption in the substrate) is
plotted as a function of temperature for eight peak frequencies ranging from 87
to 1040GHz for a 480~nm thick Nb-Si sample.  The uncertainty for the
transmission measurement is 0.02.  Inset: Transmission as a function of
frequency through a substrate with no sample (dotted line) and the sample for
two temperatures:  2.8 K (thin solid line) and 40 K (thick solid line).  The
oscillations are caused by standing waves in the sapphire substrate.
}
\label{mmtrans}
\end{figure}

\begin{figure}
\caption{${\rm Re}~\sigma(\omega)$ is plotted against the square root of the
frequency for several temperatures in the temperature range 2.8 to 32 K and
over the frequency range 87--1040 GHz.  The uncertainty in the conductivity is
$150~(\Omega {\rm m})^{-1}$. For the lowest temperatures and when $\hbar\omega
>  k_B T$, the data follow a $\omega^{1/2}$ dependence (as indicated by the
straight line). The data approach the corresponding dc values (plotted on the
vertical axis) if $\hbar\omega \ll  k_B T$.  For the highest temperatures, the
conductivity is approximately independent of frequency in this range.
}
\label{mmsigma}
\end{figure}

\begin{figure}
\caption{Log-log plot of scaled conductivity data versus scaled frequency with
the factor $C = 475~(\Omega {\rm m~K}^{1/2})^{-1}$.  The uncertainty varies
inversely with $T^{1/2}$, ranging from 20\% at 2.6 K to 6\% at 25 K.  When
temperature is 16 K and below, the data for the entire frequency range collapse
into one curve within the experimental noise.  Higher temperature data begin to
rise above the collapsed data systematically for low scaled frequencies. The
trial scaling function (dashed line) was adjusted to match approximately the
slope of the collapsed data for high frequencies but it does not describe the
data well for $\frac{\hbar \omega}{k_B T} \simeq 1$.}
\label{wtscale}
\end{figure}

%
%


\begin{references}
\bibitem{sachdev97} S. Sachdev, to appear in {\it Dynamical Properties of
Unconventional Magnetic Systems}, ed.\ A. Skjeltorp and D. Sherrington (Kluwer
Academic, Dordrecht, 1997), cond-mat/9705266.
\bibitem{damle97} K. Damle and S. Sachdev, Phys.\ Rev.\ B {\bf 56}, 8714
(1997).
\bibitem{sondhi97} S. L. Sondhi, S. M. Girvin, J. P. Carini, and D. Shahar,
Rev.\ Mod.\ Phys.\ {\bf 69}, 315 (1997) and references within.
\bibitem{engel93} L. W. Engel, D. Shahar, C. Kurdak, and D. C. Tsui, Phys.\
Rev.\ Lett.\ {\bf 71}, 2638 (1993).
\bibitem{engel95} L. W. Engel, D. Shahar, C. Kurdak, and D. C. Tsui, in {\it
Proceedings of the 11th International Conderence on High Magnetic Fields in the
Physics of Semiconductors},  p 236, ed.\ D. Heiman (World Scientific, Singapore
1995).
\bibitem{yazdani95} A. Yazdani and A. Kapitulnik, Phys.\ Rev.\ Lett.\ {\bf 74},
3037 (1995).
\bibitem{anderson58} P. W. Anderson, Phys.\ Rev.\ {\bf 109}, 1492 (1958).
\bibitem{wegner76} F. J. Wegner, Z. Physik B {\bf 25}, 327 (1976).
\bibitem{belitz94} D. Belitz and T.R. Kirkpatrick, Rev.\ Mod.\ Phys.\ {\bf 66},
261 (1994).
\bibitem{rosenbaum81} T. F. Rosenbaum, K. Andres, G. A. Thomas, and P. A. Lee,
Phys.\ Rev.\ Lett.\ {\bf 46} 568 (1981).
\bibitem{gaymann93} A. Gaymann, H. P. Geserich, and H. v. L\"ohneysen, Phys.\
Rev.\ Lett.\ {\bf 71}, 3681 (1993); Phys.\ Rev.\ B {\bf 52}, 16486 (1995).
\bibitem{song91} S. N. Song and J. B. Ketterson, Phys.\ Lett.\ A {\bf 155}, 325
(1991).
\bibitem{hertel83} G. Hertel, D. J. Bishop, E. G. Spencer, J. M. Rowell, and R.
C. Dynes, Phys.\ Rev.\ Lett.\ {\bf 50}, 743 (1983).
\bibitem{bishop85} D. J. Bishop, E. G. Spencer, and R. C. Dynes, Solid-State
Elec.\ {\bf 38}, 73 (1985).
\bibitem{allen93} L. C. Allen, M. A. Paalanen, and R. N. Bhatt, Europhys.\
Lett.\ {\bf 21}, 927 (1993).
\bibitem{carini96} J. P. Carini, H.-L. Lee, and D. V. Baxter, Ferroelectrics
{\bf 176}, 239 (1996).
\bibitem{macmillan81} W.L. McMillan, Phys.\ Rev.\ B {\bf 24}, 2739 (1981).
\bibitem{efros85} A. L. Efros and B. I. Shklovskii, in {\it Electron-Electron
Interactions in disordered systems}, ed. A. L. Efros and M. Pollak (Elsevier
1985), 409.
\end{references}
\end{document}